\documentclass[aps,pre,superscriptaddress,floatfix,twocolumn]{revtex4}
\usepackage{graphicx}
\usepackage{amsmath}
\usepackage{amsfonts}
\usepackage{amssymb}

\begin{document}
\title{Dynamics of fluctuations in a fluid below the onset of Rayleigh-B\'enard convection}
\author{Jaechul Oh}
\affiliation{Department of Physics and iQUEST, University of
California, Santa Barbara, California 93106, USA}
\author{Jos\'e M. Ortiz de Z\'arate}
\affiliation{Departamento de F\'{\i}sica Aplicada I, Facultad de
F\'{\i}sica, Universidad Complutense, 28040 Madrid, Spain}
\author{Jan V. Sengers}
\affiliation{Institute for Physical Science and Technology, and
Departments of Chemical Engineering and Mechanical Engineering, \\
University of Maryland, College Park, Maryland 20742, USA}
\author{Guenter Ahlers}
\affiliation{Department of Physics and iQUEST, University of
California, Santa Barbara, California 93106, USA}
\date{ \today}
\vskip 0.2in

\begin{abstract}

We present experimental data and their theoretical interpretation
for the decay rates of temperature fluctuations in a thin layer of
a fluid heated from below and confined between parallel horizontal
plates. The measurements were made with the mean temperature of
the layer corresponding to the critical isochore of sulfur
hexafluoride above but near the critical point where fluctuations
are exceptionally strong. They cover a wide range of temperature
gradients below the onset of Rayleigh-B\'enard convection, and
span wave numbers on both sides of the critical value for this
onset. The decay rates were determined from experimental
shadowgraph images of the fluctuations at several  camera exposure
times. We present a theoretical expression for an
exposure-time-dependent structure factor which is needed for the
data analysis. As the onset of convection is approached, the data
reveal the critical slowing-down associated with the bifurcation.
Theoretical predictions for the decay rates as a function of the
wave number and temperature gradient are presented and compared
with the experimental data. Quantitative agreement is obtained if
allowance is made for some uncertainty in the small spacing
between the plates, and when an empirical estimate is employed for
the influence of symmetric deviations from the Oberbeck-Boussinesq
approximation which are to be expected in a fluid with its density
at the mean temperature located on the critical isochore.

\end{abstract}
\pacs{PACS numbers: }

\maketitle

\section{Introduction}
\label{sec:intro}

In this paper we present experimental data for and a theoretical
analysis of the decay rates of fluctuations in a fluid layer
between two horizontal plates that are maintained at two different
temperatures. The size $L$ of the layer in the horizontal
directions is much larger than the distance  $d$
between the plates. By now it has been well established that the
presence of a uniform and stationary temperature gradient $\nabla
T_0$ in the fluid layer induces hydrodynamic fluctuations that are
long ranged in space.

An important dimensionless parameter that governs the nature of
the thermal nonequilibrium fluctuations is the Rayleigh number
\begin{equation}
R = \frac{\alpha g d^4 \nabla T_0}{D_T \nu}, \label{eq:R}
\end{equation}
where $g$ is the gravitational acceleration and where $\alpha$ is
the isobaric thermal expansion coefficient, $D_T$ the thermal
diffusivity, and $\nu$ the kinematic viscosity of the fluid. The
Rayleigh number is commonly taken to be positive when the fluid
layer is heated from below. States in which the fluid layer is
heated from above then correspond to $R < 0$. The fluid layer in
the presence of a temperature gradient remains in a quiescent
state for all $R$ less than a critical value  $R_\mathrm{c}$ for
the onset of Rayleigh-B\'enard (RB) convection, including all
negative values of $R$. The present paper is concerned with
thermal fluctuations in such a fluid layer for positive values of
$R$  below $R_\mathrm{c}$.

At sufficiently large wave numbers $q$ the intensity of
nonequilibrium fluctuations is proportional to $(\nabla
T_0)^2~q^{-4}$ both for negative and positive
$R$~\cite{KirkpatrickEtAl,LawSengers,SegreEtAl1,miPRE}. For
smaller values of $q$ it was shown
theoretically~\cite{SegreSchmitzSengers} and confirmed
experimentally~\cite{VailatiGiglio1} that the increase of the
fluctuation intensity with decreasing $q$ is quenched in the presence of
gravity, yielding a constant value in the limit of small $q$. If
the presence of top and bottom boundaries is taken into account,
one finds that the intensity of the fluctuations vanishes as
$q^2$~\cite{miPRE,Physica,Physica2} at small $q$. Hence, the
nonequilibrium structure factor $S(q)$ is predicted to exhibit a
crossover from a $q^{-4}$ dependence for larger $q$ to a $q^2$
dependence in the limit $q\to 0$, leading to a maximum at an
intermediate wave number $q_\mathrm{max}$ which has a value near
$\pi/d$. As the RB instability is approached from below, linear
theory predicts that $S(q_\mathrm{max})$, as well as the total
fluctuation power [{\em i.e.}, the integral of $S(q)$] diverges,
in agreement with the asymptotic prediction obtained by Zaitsev
and Shliomis~\cite{ZaitsevShliomis} and by Swift and
Hohenberg~\cite{SwiftHohenberg,HohenbergSwift}. The predicted
increase  as the RB instability is approached was verified
quantitatively by experiments using shadowgraph
measurements~\cite{WuEtAl}. As the fluctuations become large it
was predicted ~\cite{SwiftHohenberg,HohenbergSwift} and confirmed
experimentally~\cite{OA03} that linear theory breaks down and that
the fluctuation amplitudes saturate due to nonlinear interactions.

The present paper is concerned not with the intensity, but with
the dynamics of the nonequilibrium fluctuations. One experimental
technique to probe the time dependence of the fluctuations is
provided by dynamic Rayleigh light scattering. Dynamic light
scattering experiments in fluid layers with negative Rayleigh
numbers have shown the existence of two modes: a heat mode with a
decay rate equal to $D_T q^2$ associated with temperature
fluctuations, and a viscous mode with a decay rate equal to $\nu
q^2$ associated with transverse momentum
fluctuations~\cite{SegreEtAl1,LawEtAl}. Thus, for large $q$ where
gravity and boundary effects are negligible the coupling between
the heat mode and the viscous mode causes an enhancement of the
amplitude of nonequilibrium fluctuations, but does not affect the
decay rates, in accordance with the original predictions of
Kirkpatrick {\it et al}.~\cite{KirkpatrickEtAl}. However, for
smaller $q$ (corresponding to macroscopic wavelengths), gravity
and boundary effects induce a coupling between the decay rates of
the viscous and heat modes, as originally suggested by
Lekkerkerker and Boon~\cite{LekkerkerkerBoon}. For certain
negative values of the Rayleigh number the modes can even become
propagating~\cite{SegreSchmitzSengers,BoonEtAl,LallemandAllain,SchmitzCohen2}.
On the other hand, for positive $R$ near the RB instability the
nonequilibrium structure factor is dominated by a very slow mode
with a decay rate which vanishes as $R\to
R_\mathrm{c}$~\cite{ZaitsevShliomis,SwiftHohenberg,HohenbergSwift,LekkerkerkerBoon,KirkpatrickCohen}.

The phenomenon of critical slowing down of the fluctuations as the
RB instability is approached from below has been observed
experimentally, but the results obtained so far are
qualitative. Using forced Rayleigh scattering Allain {\it et
al}.~\cite{AllainEtAl} measured the decay of imposed horizontal
spatially periodic temperature profiles with various wave numbers
$q$. The decay time of these imposed deviations from the steady
state increased as the temperature gradient approached the value
associated with $R_\mathrm{c}$. Furthermore, the decay times
became larger for periodic temperature profiles with $q$ closer to
$q_\mathrm{c}$. Critical slowing down of nonequilibrium
fluctuations has also been observed by Sawada with an acoustic
method~\cite{Sawada}. In this experiment, first a deterministic RB
pattern was established and then the cell was turned over so as to
change the sign of $R$. The experimental observations were
interpreted with an amplitude equation that did not include any
wave-number dependence. The single decay time extracted from the
data did indeed increase as the onset of convection was
approached, but numerical agreement with the amplitude equation
was poor. Using neutron scattering, Riste and
coworkers~\cite{PedersenRiste,OtnesRiste} observed the critical
slowing down of nonequilibrium fluctuations close to the onset of
convection in a liquid crystal. Quantitative experimental studies
have been performed showing the slowing down of the dynamics of
deterministic patterns as the RB instability is approached from
{\em above}~\cite{BA77,WPDNB78}.

In the present paper we present a quantitative theoretical and
experimental study of the decay rates of fluctuations for positive
$R$ but below the RB instability. The results of both experiment
and theory reveal the critical slowing down as the RB instability
is approached from below. Theoretically, the decay rates can be
determined, in principle, from a linear stability analysis of the
deterministic Oberbeck-Boussinesq (OB) equations against
perturbations, as has been done
traditionally~\cite{LekkerkerkerBoon,Ob79,Bo03,Ch61,PlattenLegros,Manneville}.
However, here we derive these decay rates by solving the
linearized stochastic OB equations, obtained by supplementing the
deterministic OB equations with random dissipative
fluxes~\cite{ZaitsevShliomis,SwiftHohenberg,CrossHohenberg,LandauLifshitz}.
Because of Onsager's regression hypothesis, the two procedures
should yield the same result, and indeed they do. However, it is
important to note that a solution of the stochastic OB equations
is required to obtain the correct amplitudes of the nonequilibrium
fluctuations. For instance, it turns out that the wave number
$q_\mathrm{max}$ corresponding to the maximum enhancement of the
intensity of the fluctuations as a function of $q$ differs, at
least for $R<R_\mathrm{c}$, from the value of $q$ at which the
decay rate has a minimum~\cite{miPRE}.

We present in this paper the results of experimental shadowgraph
measurements~\cite{DeBruynEtAl} of the fluctuations in a thin
horizontal layer of sulfur hexafluoride heated from below but with
$R<R_\mathrm{c}$. We show that the decay rates of the fluctuations
can be obtained from an analysis of images acquired at various
exposure times, and compare the experimental results with the
theoretical predictions. The data as well as the theory reveal the
critical slowing down of the fluctuations as the onset of
convection is approached. To our knowledge, previous shadowgraph
experiments were used primarily to obtain the fluctuation
intensities~\cite{WuEtAl,BrogioliEtAl}. Exceptions are studies of
the dynamics of deterministic patterns above the RB instability
that used image sequences at constant time
intervals~\cite{MBCA96,BPA00}. Measurements of the structure
factor with the shadowgraph method can probe fluctuations at the
small wave numbers which reveal the influence of the finite
geometry. These wave numbers generally are not accessible to light
scattering.

We shall proceed as follows. In Sect.~\ref{sec:Sdynamic} we derive
the decay rates of the nonequilibrium fluctuations in a fluid
layer between two horizontal rigid boundaries. To obtain analytic
expressions we determine the dynamic structure factor in a
first-order Galerkin approximation. This section is a sequel to a
previous publication~\cite{miPRE}, in which the static
nonequilibrium structure factor was considered. In
Sect.~\ref{sec:Shadow} we then use the dynamic structure factor to
calculate the dependence of shadowgraph signals on the
exposure-time interval $\tau$. In Sect.~\ref{sec:Exper} we
describe the experimental conditions and procedures, including the
collection and analysis of the shadowgraph images. Specifically,
we show how the decay rate $\Gamma_-(q)$ of the fluctuations below
the RB instability can be determined experimentally from
shadowgraph images obtained with two different exposure times. A
comparison of the experimental results obtained for $\Gamma_-(q)$
with our theoretical predictions is presented in
Sect.~\ref{sec:Results}. Finally, our conclusions are summarized
in Sect.~\ref{sec:summary}.

\section{Dynamic structure factor and decay rates of fluctuations}
\label{sec:Sdynamic}

As was done previously~\cite{miPRE,Physica,Physica2}, we determine
the nonequilibrium structure factor by solving the linearized
stochastic OB equations for the temperature and velocity
fluctuations. It should be noted that the use of the linearized OB
equations implies that the critical slowing down of the
fluctuations near the RB instability is treated in a
mean-field approximation in which the decay rate will vanish at
$R=R_\mathrm{c}$ when the wave number $q$ equals a critical wave
number $q_\mathrm{c}$~\cite{ZaitsevShliomis}. It is known from
theory~\cite{SwiftHohenberg,GrahamPleiner} and
experiment~\cite{OA03} that nonlinear effects will cause a
saturation of the fluctuating amplitudes when they become very
large. The effect of nonlinear terms on the decay rates depends on
the nature of the bifurcation in the presence of fluctuations. In
the RB case the fluctuations change the bifurcation, which is
supercritical in the absence of fluctuations~\cite{SLB65}, to
subcritical~\cite{SwiftHohenberg,OA03}. In that case one would
expect the decay rates to remain finite at the bifurcation;
however, this issue is beyond the scope of the present paper.

For the case of a fluid layer between two plates with stress-free
({\it i.e}., slip) boundary conditions the intensity and the decay
rates of the nonequilibrium fluctuations have been obtained in a
previous publication~\cite{Physica2}. The advantage of stress-free
boundary conditions is that they permit an exact analytic solution
of the problem. Here, instead, we consider the more realistic case
of a fluid layer between two rigid boundaries, corresponding to
two perfectly conducting walls for the temperature and with
no-slip boundary conditions for the local fluid velocity. While
this case does not permit an exact analytic solution, a good
approximation can be obtained by representing the dependence of
the temperature and velocity fluctuations upon the vertical $z$
coordinate by Galerkin polynomials~\cite{Manneville}.
Specifically, two of us have determined the static nonequilibrium
structure factor in a first-order Galerkin
approximation~\cite{miPRE}. A first-order Galerkin-polynomial
solution appears to provide a good approximation below the
convection threshold for the total intensity of the
fluctuations~\cite{miPRE,EPJ}. We thus extend here the first-order
Galerkin treatment considered in Ref.~\cite{miPRE}, in the
expectation that it will also provide a good approximation for the
decay rates and the actual amplitudes of the two coupled
hydrodynamic modes.

The decay rates of the hydrodynamic modes can be readily obtained
by finding the roots in $\omega$ of the determinant of the matrix
$\mathcal{H}(\omega,q)$, defined by Eq.~(15) in Ref.~\cite{miPRE}.
There are two coupled hydrodynamic modes whose decay rates
$\Gamma_\pm(\tilde{q})$ are given by
\begin{align}\label{unoa}
&\Gamma_\pm(\tilde{q})=\frac{D_T}{2 d^2}{(\tilde{q}^2+10)}
\left\{\frac{\sigma(\tilde{q}^4+24\tilde{q}^2+504)}{(\tilde{q}^2+10)(\tilde{q}^2+12)}+1
\right. \\ &\left.
\pm\sqrt{\left[\frac{\sigma(\tilde{q}^4+24\tilde{q}^2+504)}{(\tilde{q}^2+10)(\tilde{q}^2+12)}-1\right]^2
+ \frac{27
R\sigma\tilde{q}^2}{7(\tilde{q}^2+12)(\tilde{q}^2+10)^2}}\right\},
\nonumber
\end{align}
where $\sigma\equiv \nu/\kappa$ is the Prandtl number, and
$\tilde{q}=qd$ is the dimensionless horizontal wave number of the
fluctuations. We note that in the previous
publication~\cite{miPRE} the symbol $q_\parallel$ was used to
emphasize that it represents the magnitude of a two-dimensional
wave vector $\mathbf{q}$ in the horizontal $XY$-plane. Here we
drop the subscript $\parallel$, since in the present paper the
wave vectors will always be two-dimensional in the horizontal
plane. In Eq.~(\ref{unoa}), $\Gamma_-(\tilde{q})$ represents the
decay rate of a slower heat-like mode which approaches $D_T q^2$
for large values of $q$, while $\Gamma_+(\tilde{q})$ represents
the decay rate of a faster viscous mode approaching $\nu q^2$ for
large $q$. The advantage of the Galerkin approximation is that one
can specify the decay rates $\Gamma_\pm(\tilde{q})$ explicitly as
a function of $\tilde{q}$. We note that in the first-order
Galerkin approximation considered here, we find only two decay
rates, instead of a series of decay rates that would be obtained
if higher orders were considered in the Galerkin expansion. For
studying the situation below the RB instability, where
fluctuations decay to zero, consideration of the two primary decay
rates will be adequate. As discussed more in detail later, for
these Eq.~(\ref{unoa}) is a good approximation.

We are interested in the time correlation function for the density
fluctuations at constant pressure which is directly related to the
time-dependent autocorrelation function $\langle\delta
T^*(\mathbf{q},z,t)\cdot\delta
T(\mathbf{q}^\prime,z^\prime,t^\prime)\rangle$ of the temperature
fluctuations by~\cite{miPRE}
\begin{equation}\label{defF}
\begin{split}
\langle\delta T^*(\mathbf{q},z,t)\cdot\delta
T(\mathbf{q}^\prime,z^\prime,t^\prime)\rangle= \hspace{75pt}
\\ F(q,z,z^\prime,\Delta t) \frac{(2\pi)^2}{\alpha^2\rho^2}
\delta(\mathbf{q}-\mathbf{q}^\prime).
\end{split}
\end{equation}
Here $\Delta t =|t-t^\prime|$ and $\rho$ is the average fluid
density. The time correlation function $F(q,z,z^\prime,\Delta t)$
is just the inverse frequency Fourier transform of the dynamic
structure factor $S(\omega,q,z,z^\prime)$ defined by Eq.~(9) in
Ref. ~\cite{miPRE}. Following the steps described in
Ref.~\cite{miPRE} we find after some long algebraic calculations
that the time correlation function $F(q,z,z^\prime,\Delta t)$ can
be expressed as the sum of two exponentials:
\begin{align}\label{dynamic}
&F(\tilde{q},z,z^\prime,\Delta t)=\frac{S_\mathrm{E}}{d} \left\{
\tilde{A}_+(\tilde{q})
\exp{\left[-\Gamma_+(\tilde{q})~\Delta t\right]} \right. \\
&+ \left.
\tilde{A}_-(\tilde{q})\exp{\left[-\Gamma_-(\tilde{q})~\Delta
t\right]} \right\}
\left(\frac{z}{d}-\frac{z^2}{d^2}\right)\left(\frac{z^\prime}{d}-\frac{z^{\prime
2}}{d^2}\right).\nonumber \end{align} Here the coefficient
$S_\mathrm{E}$ represents the intensity of the fluctuations of the
fluid in a local thermodynamic equilibrium state at the average
temperature $T=\bar{T}$ (see Eq.~(2) in Ref.~\cite{miPRE}). The
Galerkin polynomials appear in Eq.~(\ref{dynamic}) because they
have been used to represent the dependence of the temperature
fluctuations on the vertical coordinate $z$. We introduce the
dimensionless decay rates
\begin{equation}\label{DefTV}
\tilde{\Gamma}_\pm(\tilde{q})= t_\mathrm{v}~\Gamma_\pm(\tilde{q})
\end{equation}
where $t_\mathrm{v} \equiv d^2/D_T$ is the vertical thermal relaxation
time. The dimensionless amplitudes $\tilde{A}_\pm(\tilde{q})$ in
Eq.~(\ref{dynamic}) are then given by
\begin{widetext}
\begin{equation}\label{amplitude}
\tilde{A}_\pm(\tilde{q})=\pm 30 \frac{(\tilde{q}^2+10)\left[
\tilde{\Gamma}_\pm^2(\tilde{q})-\sigma^2 \left(
\dfrac{\tilde{q}^4+24\tilde{q}^2+504}{\tilde{q}^2+12}\right)^2
\right]-
\dfrac{27(\tilde{q}^4+24\tilde{q}^2+504)\tilde{q}^2}{28(\tilde{q}^2+12)^2}~\sigma
(\tilde{S}^0_\mathrm{NE}-\sigma R)}
{\tilde{\Gamma}_\pm(\tilde{q})~ [\tilde{\Gamma}_+^2(\tilde{q})
-\tilde{\Gamma}_-^2(\tilde{q})]}.
\end{equation}
\end{widetext}
In Eq.~(\ref{amplitude}), $\tilde{S}_\mathrm{NE}^0$ denotes the
strength of the enhancement of the static nonequilibrium structure
factor of the fluid in the absence of any boundary conditions:
\begin{equation}\label{strength}
\tilde{S}_\mathrm{NE}^0= \sigma R + \frac{(c_P/T) d^4}{D_T^2}
\left(\nabla T_0 \right)^2.
\end{equation}
Here $c_P$ is the isobaric specific heat per unit
mass~\cite{miPRE}. In Eq.~(\ref{strength}), as everywhere else in
the present paper, all thermophysical properties are to be
evaluated at the average temperature $T=\bar{T}$. We note that the
amplitudes $\tilde{A}_\pm(\tilde{q})$ depend on the Prandtl number
$\sigma$ and on the Rayleigh number $R$ not only explicitly in
accordance with Eq.~(\ref{amplitude}), but also implicitly through
expression~(\ref{unoa}) for the decay rates
$\Gamma_\pm(\tilde{q})$.

As noted in the Introduction, the decay rates of the fluctuations
can also be obtained from an analysis of deviations from steady
state on the basis of the deterministic OB equations, {\it i.e.},
the OB equations without random noise terms. Hence,
Eq.~(\ref{unoa}) for $\Gamma_\pm(\tilde{q})$ is implicit in the
standard calculation of the convection threshold within the same
Galerkin approximation employed here~\cite{Manneville}. However,
to obtain the correct amplitudes $\tilde{A}_\pm(\tilde{q})$, it is
necessary to solve the stochastic OB equations for the fluctuating
fields.

Expression~(\ref{unoa}) for the decay rates, as well as
Eq.~(\ref{amplitude}) for the amplitudes, are rather complicated.
They simplify considerably for $\sigma\to\infty$. In that limit
the decay rate of the slower mode $\tilde{\Gamma}_-(\tilde{q})$
reduces to
\begin{equation}\label{uno}
\tilde{\Gamma}_-(\tilde{q})\simeq  (\tilde{q}^2+10) -
\frac{27\tilde{q}^2 R}{28 (\tilde{q}^4+24\tilde{q}^2+504)},
\end{equation}
while the decay rate $\tilde{\Gamma}_+(\tilde{q})$ of the faster
mode becomes proportional to $\sigma$ and is so large that the
first exponential term in Eq.~(\ref{dynamic}) can be neglected.
The amplitude $\tilde{A}_-(\tilde{q})$ of the remaining
exponential term reduces, in the same $\sigma\to\infty$ limit, to
a simpler expression to be used later, in Eq.~(\ref{final}). It is
interesting to note that the limit for large Prandtl numbers is
approached rather fast. For instance, at $\sigma=15$, the
difference between the actual $\Gamma_-(\tilde{q})$ given by
Eq.~(\ref{unoa}) and the value deduced from the asymptotic
expression~(\ref{uno}) is always smaller than 3\% for any value of
$\tilde{q}$, with the larger deviations of about 3\% at
$\tilde{q}$ values close to $\tilde{q}_\mathrm{c}$. From
Eq.~(\ref{uno}) we find that $\tilde{\Gamma}_-(\tilde{q})$
approaches 10 when $\tilde{q}\to 0$, independently of the values
of the Rayleigh or Prandtl number. The value 10 obtained from our
first-order Galerkin approximation has to be compared with
$\pi^2\simeq 9.87$, obtained from the exact theory for the
limiting value at $\tilde{q}\to 0$ of this decay
rate~\cite{Manneville}.

The analysis in the previous paper~\cite{miPRE} was specifically
devoted to the static structure factor, $S(q,z,z^\prime)$, which
may be obtained by setting $\Delta t=0$ in Eq.~(\ref{dynamic}) or,
equivalently, by integrating the dynamic structure factor
$S(\omega,q,z,z^\prime)$ over the entire range of frequencies
$\omega$~\cite{miPRE}. However, as discussed in more detail in
Ref.~\cite{miPRE}, the structure factor $S(q)$ that is actually
measured in small-angle light-scattering or in
zero-collecting-time shadowgraph experiments is the one obtained
after integration of the full static structure factor over $z$ and
$z^\prime$~\cite{SchmitzCohen2}, so that:
\begin{align}\label{static}
S(\tilde{q})&=\frac{1}{d}\int_0^d dz \int_0^d dz^\prime
F(\tilde{q},z,z^\prime,0)\\&=
\frac{S_\mathrm{E}}{36}\left\{\tilde{A}_+(\tilde{q})+\tilde{A}_-(\tilde{q})\right\}=S_\mathrm{E}\left\{
\frac{5}{6}+ \tilde{S}_\mathrm{NE}^0 \Lambda_0^R(\tilde{q})
\right\},\nonumber
\end{align}
where $\Lambda^R_0(\tilde{q})$ represents the normalized
enhancement of the intensity of nonequilibrium fluctuations in the
first-order Galerkin approximation, as given by Eq.~(20) in
Ref.~\cite{miPRE}. We note that, in accordance with
Eq.~(\ref{static}), the static structure factor $S(q)$ is
expressed as the sum of an equilibrium and a nonequilibrium
contribution. However, the dynamic structure factor and its
equivalent, the time dependent correlation function given by
Eq.~(\ref{dynamic}), can no longer be written as a sum of
equilibrium and nonequilibrium contributions. This difference
between the static and the dynamic structure factor results from
the coupling between hydrodynamic modes due to the presence of
gravity and boundaries. For the same reason, the nonequilibrium
intensity enhancement $\tilde{S}^0_\mathrm{NE}$ no longer appears
as a simple multiplicative factor in the expression for the
dynamic structure factor. The observation that the nonequilibrium
dynamic structure factor is no longer the sum of an equilibrium
and a nonequilibrium contribution already pertains to the dynamic
structure factor of the ``bulk" fluid ({\it i.e}., without
considering boundary conditions), where mixing of the modes is
still caused by gravity effects~\cite{SegreSchmitzSengers}.

\section{Application to the dependence of shadowgraph signals on exposure time}
\label{sec:Shadow}

The shadowgraph method provides a powerful tool for visualizing
flow patterns in Rayleigh-B\'enard
convection~\cite{DeBruynEtAl,BPA00,TC02}. Of special interest for
the present paper is that the shadowgraph method can also be used
to measure fluctuations in quiescent nonequilibrium fluids at very
small wave
numbers~\cite{WuEtAl,VailatiGiglio1,GiglioNature,BrogioliEtAl,TC02}.

In shadowgraph experiments an extended uniform monochromatic light
source is employed to illuminate the fluid layer. Many shadowgraph
images of a plane perpendicular to the temperature gradient are
obtained with a charge-coupled-device (CCD) detector that
registers the spatial intensity distribution
$I(\mathbf{x},\tau)$ as a function of the two-dimensional position
in the imaging plane $\mathbf{x}$ and the nonzero exposure time
$\tau$ used by the detector to average photons for a single
picture. An effective shadowgraph signal ${\mathcal
I}(\mathbf{x},\tau)$ is then defined as:
\begin{equation}\label{ShadowSignal}
{\mathcal
I}(\mathbf{x},\tau)=\frac{I(\mathbf{x},\tau)-I_0(\mathbf{x},\tau)}{I_0(\mathbf{x},\tau)},
\end{equation}
where $I_0(\mathbf{x},\tau)$ is a blank intensity distribution in
the absence of any thermally excited fluctuations. In practice,
$I_0(\mathbf{x},\tau)$ is calculated as an average over many
original shadowgraph images, so that fluctuation effects cancel.

From a series of experimental shadowgraph signals ${\mathcal
I}(\mathbf{x},\tau)$, the experimental shadowgraph structure
factor $S_\mathrm{s}(\mathbf{q},\tau)$ is defined as the modulus
squared of the two-dimensional Fourier transform of the
shadowgraph signals, averaged over all the signals in the series:
$S_\mathrm{s}(\mathbf{q},\tau) \equiv \langle|{\mathcal
I}(\mathbf{q},\tau)|^2\rangle$. The physical meaning of the
shadowgraph structure factor $S_\mathrm{s}(\mathbf{q},\tau)$ in
the past has been based on the assumption that the shadowgraph
images are taken instantaneously. In the limit $\tau\to 0$
$S_\mathrm{s}(\mathbf{q},\tau)$ has been related to the static
structure factor of the fluid $S(\mathbf{q})$ by a relation of the
form~\cite{WuEtAl,BPA00,GiglioNature,TC02,miPRE}
\begin{equation}\label{Signal}
S_\mathrm{s}(\mathbf{q},0) = \mathcal{T}(q)~S(q).
\end{equation}
In Eq.~(\ref{Signal}), $\mathcal{T}(q)$ is an optical transfer
function that contains various properties, such as the wave number
of the incident light, the temperature derivative of the
refractive index of the fluid, and details of the experimental
optical arrangement. For the present work a specification of
$\mathcal{T}(q)$ is not needed since it will be eliminated in the
treatment of the experimental data in Section~\ref{sec:Exper}. The
quantity $S(q)$ in Eq.~(\ref{Signal}) equals the static structure
factor of the nonequilibrium fluid, as defined in
Eq.~(\ref{static}).

To determine the dependence of the experimental shadowgraph
structure factor $S_\mathrm{s}(q,\tau)$ on the exposure time
$\tau$ we need to extend the physical optics treatment performed
by Trainoff and Cannell~\cite{TC02}. We then find that
Eq.~(\ref{Signal}) is to be generalized to:
\begin{equation}\label{Signal2}
S_\mathrm{s}(\mathbf{q},\tau) = \mathcal{T}(q)~S(q,\tau),
\end{equation}
where $\mathcal{T}(q)$ is the same optical transfer function as in
Eq.~(\ref{Signal}) and where $S(q,\tau)$ is a new
exposure-time-dependent structure factor, which is related to the
autocorrelation function of temperature fluctuations by
\begin{align}\label{integra1}
&\frac{(2\pi)^2}{\alpha^2 \rho^2}~S(q,\tau) ~\delta(\mathbf{q}-\mathbf{q}^\prime)=\frac{1}{d \tau^2}\\
&\times \int_0^{\tau} dt \int_0^{\tau} dt^\prime \int_0^d dz
\int_0^d dz^\prime~\langle\delta T^*(\mathbf{q},z,t)\cdot\delta
T(\mathbf{q}^\prime,z^\prime,t^\prime)\rangle . \nonumber
\end{align}
\begin{figure}[t]
\includegraphics[width=7.5cm]{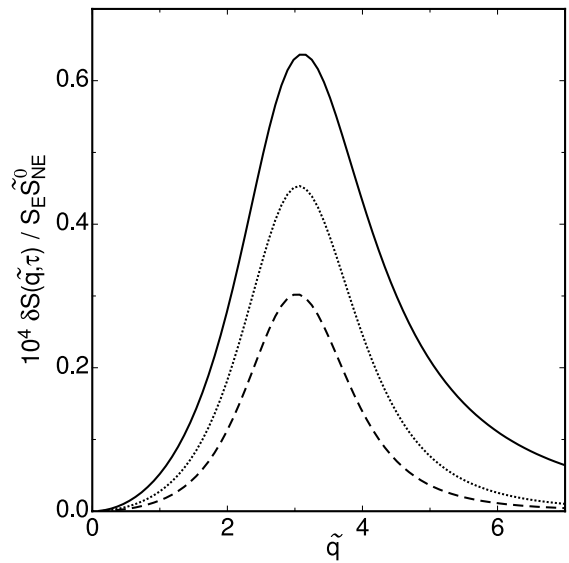}
\vskip 0.1in \caption{Difference ${\delta S(\tilde q, \tau)}/{S_\mathrm{E} \tilde S^0_\mathrm{NE}}$ (see Eq.~\ref{final} below) between the theoretical structure
factors at $R=1371$ (Eq.~(\ref{FiniteTime})), and at $R=0$
(Eq.~(\ref{dos})) as a function of $\tilde{q}$, for three
different collecting times. The solid curve corresponds to
$\tau=0$ ms, the dashed curve to $\tau=200$ ms, and the dotted
curve to $\tau=500$ ms. The Prandtl number is $\sigma=34$ and
$t_\mathrm{v} = 0.74$ s.} \label{f:uno}
\end{figure}
In principle, $S(q,\tau)$ can be evaluated by substituting
Eq.~(\ref{defF}) with Eq.~(\ref{dynamic}) for the correlation
function of the density fluctuations into the right-hand side of
Eq.~(\ref{integra1}). However, for the experimental results to be
presented, it is sufficient to evaluate $S(q,\tau)$ for large
values of the Prandtl number $\sigma$, in which case the
exponential contribution with decay rate $\Gamma_+(q)$ in
Eq.~(\ref{dynamic}) can be neglected, as discussed before.
Retaining only the contribution from the slower mode with
amplitude $\tilde{A}_-(q)$ in Eq.~(\ref{dynamic}) and performing
the integrations, we deduce from Eq.~(\ref{integra1})
\begin{equation}\label{FiniteTime}
S(\tilde{q},\tilde{\tau})=S_\mathrm{E} \tilde{A}_-(\tilde{q})
\frac{\tilde{\tau}~\tilde{\Gamma}_-(\tilde{q}) -1+
\exp[-\tilde{\tau}~\tilde{\Gamma}_-(\tilde{q})]} {18
\tilde{\tau}^2~ \tilde{\Gamma}^2_-(\tilde{q})},
\end{equation}
where $\tilde{\tau}$ is a dimensionless exposure time, defined by
$\tilde{\tau}=\tau/t_\mathrm{v}$ with $t_\mathrm{v}$ the
vertical relaxation time introduced in Eq.~(\ref{DefTV}). In the
limit $\tau\to 0$, Eq.~(\ref{FiniteTime}) reduces to
$S(\tilde{q})=S_\mathrm{E} \tilde{A}_-(\tilde{q})/36$, so that it
equals the static structure factor measured in small-angle
light-scattering or in shadowgraph experiments in the zero
collecting time approximation, as given by Eq.~(\ref{static}). It
is interesting to note that, due to the nonzero collecting time,
$\tau$, even in thermal equilibrium ($R=0$) the shadowgraph
measurements present some ``structure". From
Eq.~(\ref{FiniteTime}), we find that this equilibrium structure,
as a function of the dimensionless collecting time,
$\tilde{\tau}$, is given by
\begin{equation}\label{dos}
S_\mathrm{E}(\tilde{q},\tilde{\tau}) = S_\mathrm{E}
\frac{5}{3\tilde{\tau}^2}
\frac{\tilde{\tau}(\tilde{q}^2+10)-1+\exp[-\tilde{\tau}(\tilde{q}^2+10)]}{(\tilde{q}^2+10)^2}.
\end{equation}
It can be readily checked that, in the limit $\tau\to 0$,
Eq.~(\ref{dos}) reduces to the structureless constant
$(5/6)S_\mathrm{E}$, in agreement with Eq.~(\ref{static}). This
value is 17\% lower than the actual value $S_\mathrm{E}$, due to
the use of a first-order Galerkin approximation~\cite{miPRE}.

To gain insight into the effect of the exposure
time on shadowgraph measurements, we show in Fig.~\ref{f:uno} the
difference between the nonequilibrium structure factor 
 as obtained from Eq.~(\ref{FiniteTime}) and the
equilibrium structure factor ($R=0$) as obtained from
Eq.~(\ref{dos}), as a function of $\tilde{q}$, for three different
collecting times. Corresponding to some of the experimental results to be 
discussed below, we used $R=1371$, a vertical relaxation time
$t_\mathrm{v}=d^2/D_T$ of 0.74 s,  and a Prandtl number
$\sigma$ equal to 34. We evaluated the difference for 
$\tau =0$, $\tau= 200$, and $\tau=500$ ms. 
The structure factors $S(q,\tau)$ and
$S_\mathrm{E}(q,\tau)$ have been normalized by dividing them by
the product $S_\mathrm{E}\tilde{S}_\mathrm{NE}^0$.

From Fig.~\ref{f:uno}, we arrive at the following conclusions: (i)
The main effect of a nonzero collecting time is to lower the
height of the measured $S(q,\tau)$; this is expected since
fluctuations cancel out when larger exposure times are used. (ii)
An additional effect of a nonzero collecting time, which can be
observed in Fig.~\ref{f:uno}, is a displacement of the maximum in
$S(q,\tau)$ to lower $\tilde{q}$ values. This effect is mainly due
to the subtraction of the equilibrium structure given by
Eq.~(\ref{dos}), which decreases with increasing $\tilde{q}$.
(iii) Another interesting feature, we infer from Fig.~\ref{f:uno},
is that the dependence on $\tilde{q}^2$ for low $\tilde{q}$ is
preserved, while the dependence on $\tilde{q}^{-4}$ for large
$\tilde{q}$ is destroyed for the effective structure factor.
Actually, at large $q$, there exists a crossover from a $q^{-4}$
dependence to a $q^{-6}$ dependence, for nonzero values of 
 $\tau$, as will be shown below in Eq.~(\ref{G32}).

A simpler formula for $S(\tilde{q},\tilde{\tau})$ can be obtained
by introducing into Eq.~(\ref{FiniteTime}) the limiting value of
$\tilde{A}_-(q)$ for large Prandtl numbers. We then obtain
\begin{multline}\label{final}
\frac{\delta S(\tilde q, \tau)}{S_\mathrm{E} \tilde S^0_\mathrm{NE}} = \frac{S(\tilde{q},\tilde{\tau})-S_\mathrm{E}(\tilde{q},\tilde{\tau})}{S_\mathrm{E}
\tilde{S}^0_\mathrm{NE}}\\ = \frac{5}{3\sigma}
\frac{\tilde{\tau}\tilde{\Gamma}_-(\tilde{q}) -1+
\exp[-\tilde{\tau}\tilde{\Gamma}_-(\tilde{q})]}
{\tau^2~\tilde{\Gamma}^2_-(\tilde{q})}\\ \hfill\times  \frac{27
\tilde{q}^2}{28 (\tilde{q}^2+10)(\tilde{q}^4 +
24\tilde{q}^2+504)-27 R \tilde{q}^2},
\end{multline}
where for $\tilde{\Gamma}_-(\tilde{q})$ Eq.~(\ref{uno}) for large
$\sigma$ should be used. As discussed after Eq.~(\ref{uno}), for
Prandtl numbers near 15 the asymptotic
limit~(\ref{final}) is already closely approached. Equation~(\ref{final})
reduces, in the limit $\tau\to 0$, to the expressions obtained in
Ref.~\cite{miPRE}. It is interesting to note that, in the limit
$\tilde{q}\to\infty$, Eq.~(\ref{final}) reduces to:
\begin{equation}
\frac{\delta S(\tilde q, \tau)}{S_\mathrm{E} \tilde S^0_\mathrm{NE}} \xrightarrow{\tilde{q}\to \infty}~
\frac{45}{28\tilde{\tau}\sigma}~\frac{1}{\tilde{q}^6}, \label{G32}
\end{equation}
showing, as mentioned above, that a nonzero collecting time
changes the asymptotic behavior from $q^{-4}$ to $q^{-6}$. Note
that Eq.~(\ref{G32}) is evidently not valid when $\tau=0$ in which
case one needs the asymptotic expression given by Eq.~(30) in
Ref.~\cite{miPRE}.

\section{Experiments}
\label{sec:Exper}

\subsection{Apparatus}
\label{sec:apparatus}

To visualize the thermal fluctuations with the shadowgraph method
one needs to perform the experiments with a fluid in which the
thermal noise will be
large~\cite{HohenbergSwift,VanBeijerenCohen}. This goal can be
accomplished by selecting a fluid in the vicinity of its critical
point~\cite{WuEtAl,DeBruynEtAl,OA03}. The measurements reported
here were obtained for sulfur hexafluoride. The apparatus and
experimental procedures have been described in detail
elsewhere~\cite{DeBruynEtAl}. Here we describe only those aspects
which are specific to the experiment with a fluid near its
critical point.

For the details of the cell construction we refer to Fig. 9 of
Ref.~\cite{DeBruynEtAl}. Initially we used a diamond-machined
aluminum bottom plate which could be positioned with
piezo-electric elements. The bottom-plate thermistors were
embedded approximately 0.64 cm below its top surface. Even though
to the naked eye this plate had a near-perfect mirror finish, the
tool marks from the diamond machining imposed a preferred
direction on the fluctuations below the onset of convection. Thus
we placed an optically flat sapphire of thickness 0.318 cm on top
of the aluminum plate. A thin silver film was evaporated on the
top surface of this sapphire to provide a mirror for the
shadowgraphy.

Initially we used a sapphire of thickness 0.952 cm for the cell
top. In addition a sapphire of thickness 1.90 cm was in the optical
path of the shadowgraphy and  provided the top window of the
pressure vessel. In combination with the very small cell thickness
used in these experiments, it turned out that the optical
anisotropy of these randomly oriented sapphires introduced an
anisotropy of the shadowgraph images which obscured the rotational
symmetry of the fluctuations. In order to minimize this effect, we
replaced the pressure window by a fused-quartz window and used an
optically flat sapphire of 0.318 cm thickness for the cell top.
Under these conditions we found that the fluctuations below the
onset of convection yielded a structure factor which was nearly
invariant under rotation, as can be seen from Fig.~\ref{fig:images} below.

The cell spacing was fixed by a porous paper sidewall with an
inner (outer) diameter  of 2.5 (3.5) cm. Since the top sapphire
was supported along its perimeter which had a diameter of 10 cm
({\it i.e.}, considerably larger than the cell wall), the force
exerted on the cell top by the  bottom plate and the wall caused a
slight bowing of the initially flat top. Over the entire sample
diameter this yielded a radial cell-spacing variation
corresponding to about one circular fringe when illuminated with
an expanded parallel He-Ne laser beam. This variation of the
thickness by about 0.3 $\mu$m assured that convection would start
in the cell center, rather than being nucleated inhomogeneously
near the cell wall. Assuming a parabolic radial profile for the
cell spacing, we estimate that the spacing was uniform to much
better than 0.1\% over the 1.3$\times$1.3 cm$^2$ area near the
cell center which was actually used for the shadowgraph images.
The actual sample thickness was measured
interferometrically~\cite{DeBruynEtAl} and found to be 34.3
$\mu$m.

\subsection{Properties of SF$_6$ near the critical point}
\label{sec:properties}

The thermodynamic properties of sulfur hexafluoride in the
critical region can be calculated from a equation developed by
Wyczalkowska and Sengers~\cite{WS99}. For the viscosity we used a
fit of data from Refs.~\cite {HBT85} and \cite{SV89} to a smooth
function. This approach neglects a small anomaly of the viscosity
at the critical point. We used fits of smooth functions to the
conductivity data from
Refs.~\cite{LK65,SH73,LimSwinney,Li73,BT81,KI85}.

\begin{figure}[t]
\includegraphics[width=7cm]{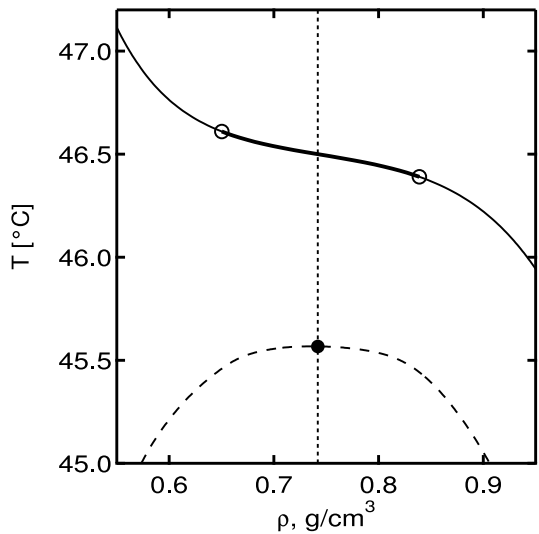}
\vskip 0.1in \caption{The temperature-density plane near the
critical point of SF$_6$. The dashed line indicates the
coexistence curve separating liquid and vapor. The vertical dotted
line is the critical isochore. The solid circle is the critical
point with $T_\mathrm{c} = 45.567^\circ$C, $P_\mathrm{c} = 37.545$
bar, $\rho_\mathrm{c} = 0.742$ g/cm$^3$. The solid line represents
the isobar $P = 38.325$ bar used in our measurements, and the
heavy section of this line shows the temperature-density range on
the isobar which is spanned by the sample with thickness $d =
34.3~\mu$m when $\Delta T = 0.22 ^\circ$C.} \label{fig:phase_dia}
\end{figure}

The measurements were made at constant pressure $P$ and at
constant mean sample temperature $\bar T$. The mean  temperature
was adjusted so that the density $\rho(\bar T)$ was the critical
density $\rho_\mathrm{c}=742$ kg m$^{-3}$. The imposed temperature
difference $\Delta T$ caused a density variation of the sample
along an isobar. This is illustrated in Fig.~\ref{fig:phase_dia}
for the conditions of the present experiments, namely for $P =
38.325$ bar and $\bar T = 46.50 ^\circ$C. At
$\rho=\rho_\mathrm{c}$, we have $\sigma = 34.0$, $D_T = 1.59\times
10^{-5}$ cm$^2$/s, and $t_\mathrm{v} \equiv d^2/D_T = 0.738$ s.

\begin{figure}[t]
\includegraphics[width=7.5cm]{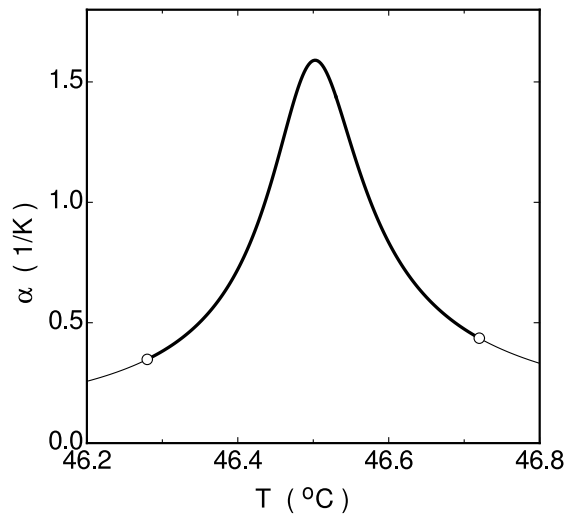}
\caption{The thermal expansion coefficient $\alpha$ along the
isobar $P = 38.325$ bars.  The heavy part of the line terminated
by two open circles indicates the temperature range spanned by the
sample with $d = 34.3 ~\mu$m and $\Delta T = \Delta T_\mathrm{c} =
0.44^\circ$C.} \label{fig:alpha}
\end{figure}

\subsection{Symmetric departures from the Oberbeck-Boussinesq approximation}
\label{sec:NOB}

Much of the theoretical work on RBC has been done in the OB
approximation~\cite{Bo03,Ob79}, which assumes that the fluid
properties do not vary over the imposed temperature interval,
except for the density where it provides the buoyant
force~\cite{Ch61}.  Non-OB effects have been considered by a
number of investigators, and most systematically by
Busse~\cite{Bu67a}. At leading order, they break the reflection
symmetry of the system about the horizontal midplane, and at the
onset of convection they yield a transcritical bifurcation to a
hexagonal pattern~\cite{BPA00}, instead of the roll pattern of
pure OB convection. When the mean temperature corresponds to the
critical isochore, $\rho=\rho_\mathrm{c}$, this effect is of
modest size. Although several properties contribute, we illustrate
this by showing the isobaric thermal expansion coefficient in
Fig.~\ref{fig:alpha} along the isobar of Fig.~\ref{fig:phase_dia}
as an example. One can approximate $\alpha$ as a sum of two
contributions, one of which is anti-symmetric and the other one
symmetric about the mean temperature (and thus approximately also
about the horizontal midplane) of the sample. Only the
anti-symmetric part is considered in the theory~\cite{Bu67a}. Its
smallness near the onset of convection is seen from the similar
values of $\alpha$ at the top and bottom of the cell (open circles
in the figure). A quantitative calculation of the parameter $\cal
P$ introduced by Busse \cite{Bu67a} (see Eqs.~(13) in
Ref.~\cite{BPA00}) yields ${\cal P} = -0.23$. This value indicates
that, within our experimental resolution, only rolls should be
seen near onset. This is indeed the case~\cite{OA03}. However, the
variation of $\alpha$ which is symmetric about the midplane is
quite large. It does not break the reflection symmetry about that
plane and thus permits the existence of {\it rolls} at onset
rather than requiring a hexagonal plan form. Behavior similar to
that of $\alpha$ is found for the specific heat $c_P$ and for the
Rayleigh number. At present there is no theoretical treatment of
these higher-order non-OB effects. Thus we proceed empirically by
examining the variation of $R$ as a function of vertical position
or local sample temperature. If we neglect the temperature
dependence of the thermal conductivity and assume that the local
temperature in the fluid layer still varies linearly as a function
of $z$, we obtain the Rayleigh number profiles shown in
Fig.~\ref{fig:Rayleigh}. The curve at the top is for the
experimental value $\Delta T = \Delta T_\mathrm{c} = 0.44^\circ$C
corresponding to the onset of convection. The other curves, from
top to bottom, correspond to $\Delta T/\Delta T_\mathrm{c} =
$0.860, 0.699, 0.430, and 0.269. For $\Delta T = 0.44 ^\circ$C one
sees that the local $R(\rho_\mathrm{c}) \simeq 3190$ far exceeds
the value $R_\mathrm{c} = 1708$ for the uniform system. On the
other hand, near the top and bottom of the sample the local $R$ is
well below the onset of convection for the homogeneous system. The
data in Fig.~\ref{fig:Rayleigh} show that, though we find rolls
above threshold, non-OB effects cannot be completely neglected in
our experiments. We shall return to the influence of non-OB
effects on the comparison between experiment and theory in
Sect.~\ref{sec:Results}.

\begin{figure}[t]
\includegraphics[width=7.5cm]{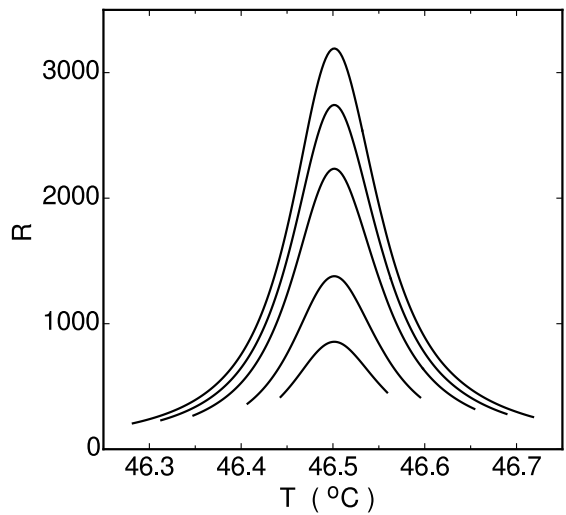}
\vskip 0.1in \caption{The Rayleigh number $R$ (see
Eq.~(\ref{eq:R})) along the isobar $P = 38.325$ bar for $d = 34.3
~\mu$m. From top to bottom, the curves are for $\Delta T/\Delta
T_\mathrm{c}  = 1.000$, 0.860, 0.699, 0.430, and 0.269 with
$\Delta T_\mathrm{c} = 0.44^\circ$C.} \label{fig:Rayleigh}
\end{figure}

\subsection{Sample temperature}
\label{sec:tbar}

The bath temperature $T_\mathrm{bath}$ and the bottom-plate
temperature $T_\mathrm{BP}$ were adjusted so as to hold the mean
sample temperature constant.  This temperature was chosen so that
the mean density corresponded to the critical density
$\rho_\mathrm{c}$. Because of the small sample thickness the
thermal resistance of the sample was comparable to that of the top
and bottom confining plates. This fact required a special
procedure to assure that the sample was indeed at the temperature
corresponding to $\rho_\mathrm{c}$. Before a run at a given fixed
pressure was started, we measured the power of shadowgraph images
of the fluctuations at a fixed imposed $\Delta T_\mathrm{ext} =
(T_\mathrm{BP} - T_\mathrm{bath})$ as a function of the mean
temperature $\bar T_\mathrm{ext} = (T_\mathrm{BP} +
T_\mathrm{bath})/2$ of the system consisting of the bottom plates,
the sample, and the top plate as determined by the bath
temperature and the bottom-plate thermistor. This temperature
difference is the sum of those across the bottom aluminum plate
$\Delta T_\mathrm{Al}$, across the boundary between the aluminum
plate and the bottom sapphire $\Delta T_\mathrm{b1}$, across the
bottom sapphire $\Delta T_\mathrm{sb}$, across the sample  $\Delta
T$, across the top sapphire $\Delta T_{st}$, and across a boundary
layer above the top sapphire in the water bath $\Delta
T_\mathrm{b2}$. From estimates of the thermal resistances of these
sections we find $\Delta T/\Delta T_\mathrm{ext} = 0.473$. Thus,
at the onset of convection, we measured $\Delta T_\mathrm{ext,c} =
0.930^\circ$C and deduced $\Delta T_\mathrm{c} = 0.44^\circ$C.

\begin{figure}[t]
\includegraphics[width=7.5cm]{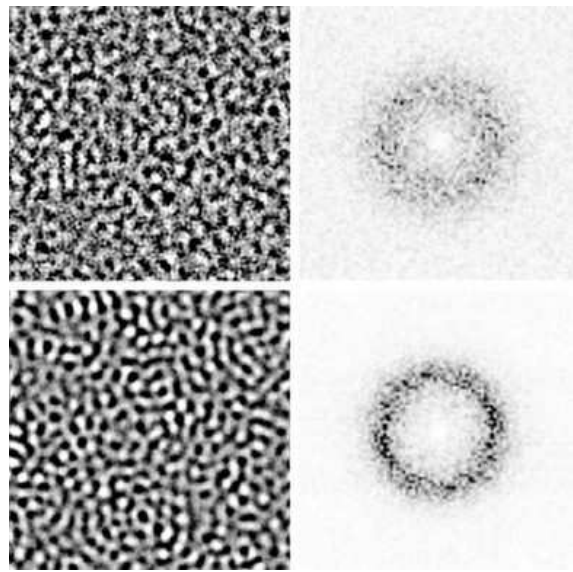}
\vskip 0.1in \caption{Shadowgraph signals (left column) and the moduli squared of their Fourier transforms (right colum) for $\Delta T = 0.189$ (top row) and $\Delta T = 0.378$ K (bottom row). The exposure time was 500 ms.} \label{fig:Rayleigh}
\label{fig:images}
\end{figure}

\subsection{Analysis of shadowgraph images}
\label{sec:analysis}

At each $\Delta T$, three shadowgraph-image sequences $I_i({\bf
x},\tau_j),~i = 1, ..., N$ with $N = 1024$, and $j = 0,1,2$ were
acquired. For each sequence, a different exposure time $\tau_j$
was used, namely $\tau_0$ = 0.500 s, $\tau_1$ = 0.350 s, and
$\tau_2$ = 0.200 s. The time interval $\delta t$ between the
images was typically one or two seconds, which was large enough
for the images to be nearly uncorrelated. The images of each
sequence were averaged to provide a background image $I _0({\bf
x},\tau_j)$, as discussed in Sect.~\ref{sec:Shadow}. Then, for
each image of the sequence, a dimensionless shadowgraph signal
$\mathcal{I}_i({\bf x},\tau_j)$ was computed, in accordance with
Eq.~(\ref{ShadowSignal}). The mean (over $\bf x$) value of a
typical ${\cal I}_i(\mathbf{x},\tau_j)$ was within the range $\pm
0.01$, indicating adequate stability of the light intensity and
image-acquisition system. Next, the two-dimensional Fourier
transform of each shadowgraph signal in the sequence was computed,
and the modulus square calculated, obtaining series of
$|\mathcal{I}_i(\mathbf{q},\tau_j)|^2$ for further analysis.
Typical shadowgraph signals and the squares of the moduli of 
their Fourier transforms are shown in Fig.~\ref{fig:images}. 
The nearly-uniform angular distribution of the transforms illustrates
 the rotational invariance of the Rayleigh-B\'enard system.

As explained in Sect.~\ref{sec:Shadow}, due to the rotational
invariance of RB convection in the horizontal plane, the modulus
squared Fourier transformed shadowgraph signals have rotational
symmetry, and for an infinitely extended sample they would depend
only on the modulus $q$ of the wave vector $\mathbf{q}$. However,
the finite spatial extent of the images leads to random angular
fluctuations of $|\mathcal{I}_i(\mathbf{q},\tau_j)|^2$. To reduce
these fluctuations, we performed azimuthal averages
$\overline{|\mathcal{I}_i(\mathbf{q},\tau_j)|^2}$ over thin rings
in Fourier space (the angular average is denoted by the overline
and depends only on $q$ and $\tau_j$). Now for each shadowgraph measurement
$I_i(\mathbf{x},\tau_j)$ the integral
\begin{equation}
P_i(\tau_j) = \int_0^\infty 2\pi
q~\overline{|\mathcal{I}_i(\mathbf{q},\tau_j)|^2}~dq \label{eq:P}
\end{equation}
is the total power and, by Parseval's theorem, has to be equal to
the variance of the original $\mathcal{I}_i({\bf x},\tau_j)$. We
used Eq.~(\ref{eq:P}) as a check of consistency for the entire
procedure of taking the Fourier transform, of calculating the
modulus squared and the azimuthal average over thin rings, and of
assigning to each ring a $q$ value. Finally, we averaged over the $N$ individual
$\overline{|\mathcal{I}_i(\mathbf{q},\tau_j)|^2}$ of each $\tau_j$
series, to obtain the experimental shadowgraph structure factor
$S_\mathrm{s}(q,\tau_j) \equiv
\langle\overline{|\mathcal{I}_i(\mathbf{q},\tau_j)|^2}\rangle$.
\begin{figure}[t]
 \includegraphics[width=7.5cm]{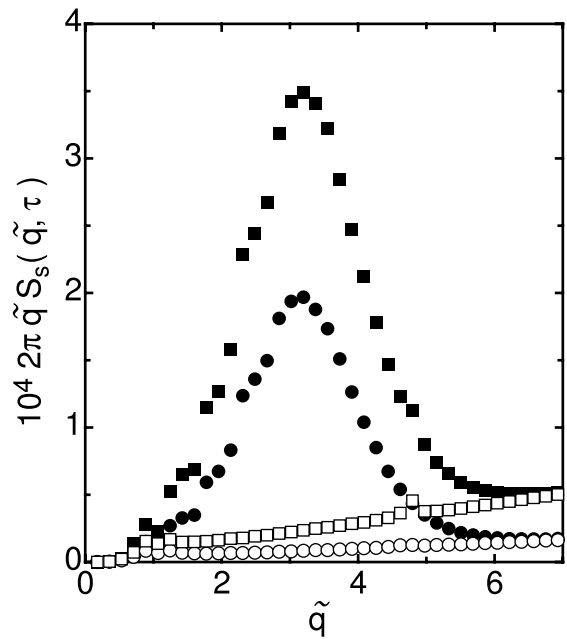}
\vskip 0.1in \caption{Experimental shadowgraph structure factor
$2\pi \tilde{q} S_\mathrm{s}(\tilde{q},\tau)$ as a function of the
dimensionless wave number $\tilde{q}$. These results are for $P =
38.325$ bar, $\bar T =46.5^\circ $C, and $\Delta T =
0.189^\circ$C. The solid squares (solid circles) are for $\tau =
0.200$ s ($\tau = 0.500$ s). The open symbols are the
corresponding background measurements for $\Delta T = 0$.}
\label{fig:S_s}
\end{figure}

For the remainder of this paper we shall use the experimentally 
determined cell spacing $d = 34.3 \mu$m to scale $q$ 
according to $\tilde q = q d$.
Two examples of the product $2 \pi\tilde{q}
S_\mathrm{s}(\tilde{q},\tau_j)$, both for $P = 38.325$ bar, $\bar
T = 46.5^\circ$C, and $\Delta T = 0.189^\circ$C, are shown in
Fig.~\ref{fig:S_s}. The solid squares are for $\tau_2 = 0.200$ s,
whereas the solid circles are for $\tau_0 = 0.500$ s. As discussed
in the previous section, the relationship~(\ref{Signal2}) between
the experimental shadowgraph structure factor
$S_\mathrm{s}(\tilde{q},\tau_j)$ and the corresponding fluid structure
factor $S(\tilde{q},\tau_j)$  involves the optical transfer function
$\mathcal{T}(\tilde{q})$. Nevertheless, we observe some qualitative
agreement of the experimental results for $2\pi \tilde{q}
S_\mathrm{s}(\tilde{q},\tau)$ in Fig.~\ref{fig:S_s} and the theoretical
results for $S(\tilde{q},\tau)$ shown in Fig.~\ref{f:uno}. The expected
maximum near $\tilde{q} = \tilde{q}_c$ is present, and the decrease as $\tilde{q}$
vanishes is consistent with the predicted $\tilde{q}^2$ dependence. The
longer averaging of the random fluctuations diminishes $2\pi
\tilde{q}~S_\mathrm{s}(\tilde{q},\tau)$, which was one of the features noted after
Fig.~\ref{f:uno} for $S(\tilde{q},\tau)$.

The experimental structure factor $S_\mathrm{s}(\tilde{q},\tau_j,\Delta
T)$ depends on the Rayleigh number and, hence, on $\Delta T$. For
$\Delta T = 0$, instrumental (mostly camera) noise is expected to
be the dominant contribution, {\it i.e}., the contribution from
the equilibrium fluctuations is negligible and the theoretical
equilibrium structure, given by Eq.~(\ref{dos}), is expected to be
unobservable in our experiments. Examples of the shadowgraph structure factors for
$\Delta T = 0$ are shown by the open symbols in
Fig.~\ref{fig:S_s}. One sees that $2\pi \tilde{q} S_s(\tilde{q},\tau_j,0)$ is well
represented by a straight line, corresponding to white noise. At large $\tilde{q}$, $2\pi \tilde{q}
S_s(\tilde{q},\tau_j,0)$ merges smoothly into the data for the $2\pi \tilde{q}
S_s(\tilde{q},\tau_j,\Delta T)$ with the same $\tau_j$, as one would
expect.
\begin{figure}[t]
 \includegraphics[width=7.5cm]{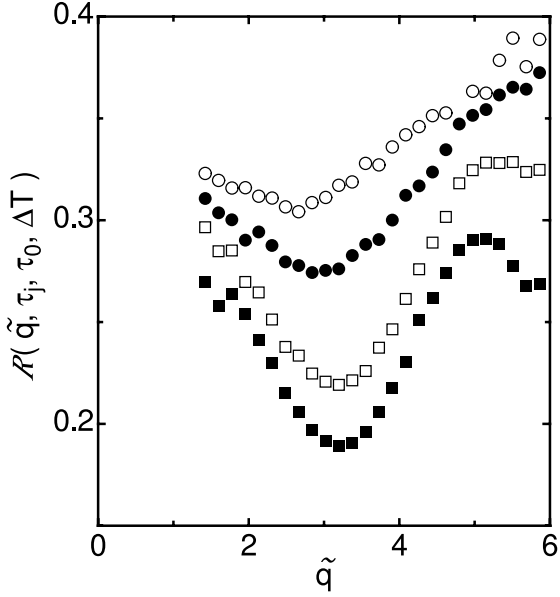}
\vskip 0.1in \caption{The ratio ${\cal R}(\tilde q,\tau_j,\tau_0,\Delta
T)$ (see Eq.~(\ref{eq:ratio})) based on $\tau_j = 0.200$ s and
$\tau_0 = 0.500$ s. From top to bottom the data are for $\Delta T
= 0.118,~0.189,~0.307$ and $0.378^\circ$C. The onset of convection
occurred at $\Delta T_\mathrm{c} = 0.440^\circ$C.}
\label{fig:ratio}
\end{figure}

A detailed comparison between the experimental and the theoretical
structure factors requires knowledge of the optical transfer
function, $\mathcal{T}(\tilde{q})$, which depends on the details of the
experimental optical arrangement. It involves, for instance, the
size of the pinhole and the focal length of the lens used to make
the ``parallel" beam, and the spectral width of the light source.\cite{TC02}
In the present work the spatial structures to be determined (the
fluctuation wavelengths) had length scales [typically ${\cal
O}(50~\mu {\rm m})$] that are one or two orders of magnitude
smaller than those of more conventional RB experiments. For this
reason, we found it difficult to obtain ${\cal T}(\tilde{q})$ for our
instrument with sufficient accuracy to avoid significant
distortion of $S(\tilde{q})$. We circumvented the difficulty of the
optical transfer function by deriving the {\it dynamic} properties
of the fluctuations from ratios of $S_\mathrm{s}(\tilde{q},\tau_j)$ with
different values of $\tau_j$. To account for the instrumental
white noise, before taking such ratios, we subtracted the measured
$S_\mathrm{s}(\tilde{q},\tau_j,0)$ in the absence of a temperature
gradient from $S_\mathrm{s}(\tilde{q},\tau_j,\Delta T)$ to yield
\begin{equation}
\delta S_\mathrm{s}(\tilde{q},\tau_j,\Delta T) = S_s(\tilde{q},\tau_j,\Delta T) -
S_\mathrm{s}(\tilde{q},\tau_j,0).
\end{equation}
After this background subtraction, we formed for $j=1,2$ the ratio
\begin{equation}\label{eq:ratio}
{\cal R}(\tilde{q},\tau_j,\tau_0,\Delta T) \equiv \frac{\tau_j^2~ \delta
S_\mathrm{s}(\tilde{q},\tau_j,\Delta T)}{\tau_0^2~\delta
S_\mathrm{s}(\tilde{q},\tau_0,\Delta T)}\ .
\end{equation}
where $\tau_0$ is the longest exposure time $\tau_0 = 0.500$ s.
The ratios $\mathcal{R}$ obtained for $\tau_j = 0.200$ s as a
function of $\tilde{q}$ are shown in Fig.~\ref{fig:ratio} for four
values of the temperature difference $\Delta T$. The shadowgraph
transfer function ${\cal T}(\tilde{q})$ cancels and is no longer contained
in ${\cal R}$. In addition, for the ratios $\mathcal{R}$, it is
irrelevant whether we use the definition of the structure factor
considered in Sect.~\ref{sec:Sdynamic}, or the shadowgraph
definition displayed in Fig.~\ref{fig:S_s}, the latter including a
factor $2\pi \tilde{q}$. Thus, we are allowed to use
Eq.~(\ref{FiniteTime}) for $\delta S_\mathrm{s}$, so that
\begin{equation}
{\cal R}(\tilde{q},\tau_j,\tau_0,\Delta T) = \frac{\tau_j\Gamma(\tilde{q}) - 1 +
\exp(-\tau_j\Gamma(\tilde{q}))}{\tau_0\Gamma(\tilde{q}) - 1 +
\exp(-\tau_0\Gamma(\tilde{q}))}\ . \label{eq:calR}
\end{equation}
One sees that ${\cal R}$ depends only on $\tau_0, \tau_j, \tilde{q},$ and
$\Gamma(\tilde{q},\Delta T)$. At each $\tilde{q}$ and $\Delta T$, only the decay
rate $\Gamma$ is unknown and thus can be determined from the
experimental value of $\cal R$.

\begin{figure}[t]
 \includegraphics[width=7.5cm]{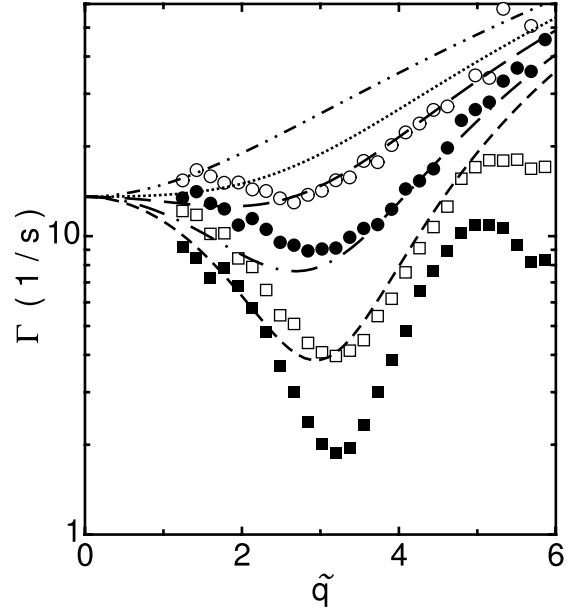}
\vskip 0.1in \caption{Results for $\Gamma(\tilde q,\Delta T)$, in units
of inverse seconds, obtained from the values of $\cal R$ shown in
Fig.~\ref{fig:ratio} (see Eq.~(\ref{eq:calR})). The symbols are
the same as in Fig.~\ref{fig:ratio}. The top (dash-double-dotted)
curve is the prediction for $\Delta T = 0$. The remaining curves
are the theoretical predictions (see Eq.~(\ref{uno})) for the
values of $\Delta T$ of the data (see Fig.~\ref{fig:ratio}).}
\label{fig:Gamma0}
\end{figure}

\begin{figure}[t]
 \includegraphics[width=8cm]{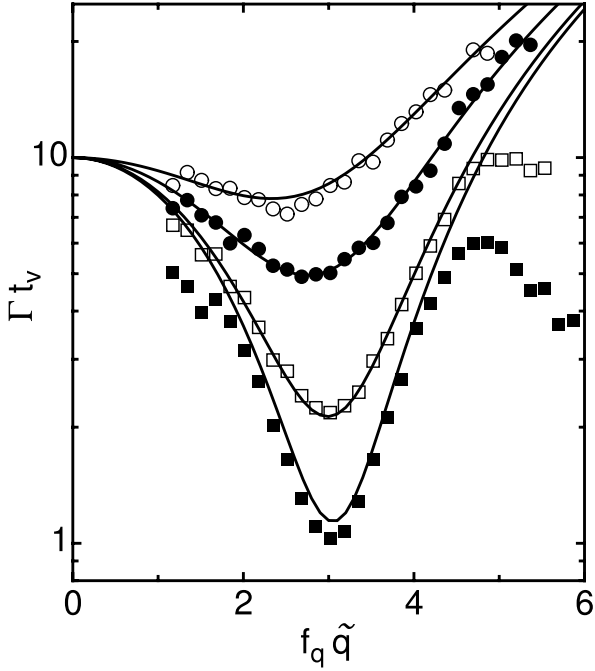}
\vskip 0.1in \caption{Results for $\Gamma t_\mathrm{v}$ as a
function of $f_q \tilde{q}$, using the values $t_\mathrm{v} =
0.551$ s and $f_q = 0.944$ from the least-squares fit described in
the text. The data and symbols correspond to those in
Fig.~\ref{fig:Gamma0}. The curves indicate the corresponding
theoretical results obtained from Eq.~(\ref{uno}) by using the
scale factors of the Rayleigh number $f_{R,k}$ from the
least-squares fit.} \label{fig:Gamma1}
\end{figure}

\begin{figure}[t]
 \includegraphics[width=7.5cm]{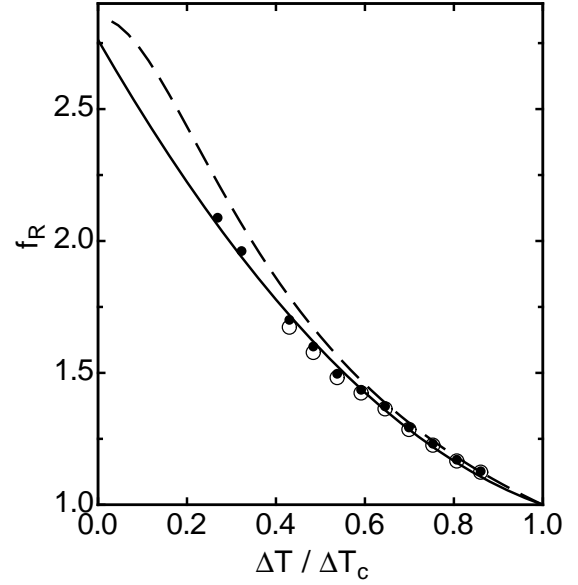}
\vskip 0.1in \caption{Values of the fit parameter $f_{R,k}$ obtained
from the least-squares fit described in the text. Solid symbols
correspond to $\Gamma$ based on exposure times $\tau_0 = 0.500$ s
and $\tau_j = 0.200$ s. For the open symbols $\tau_j$ was 0.350 s.
The solid curve is the polynomial $f_{R} = 2.74 - 2.87 \Delta
T/\Delta T_\mathrm{c} + 1.13(\Delta T/\Delta T_\mathrm{c})^2$
which passes through the point (1,1). The dashed curve represents
$\langle R \rangle / \langle R \rangle_\mathrm{c}$ and was
calculated from values of $R$ like those shown in
Fig.~\ref{fig:Rayleigh}.} \label{fig:paras}
\end{figure}

\section{Experimental results and comparison with theory}
\label{sec:Results}

In Fig.~\ref{fig:Gamma0} we show the decay rate $\Gamma$ as a
function of $\tilde{q}$. The symbols represent the experimental
values deduced from the data for $\mathcal{R}$ displayed in
Fig.~\ref{fig:ratio} by solving Eq.~(\ref{eq:calR}). From top to
bottom, the data sets are for $\Delta T/\Delta T_\mathrm{c} =$
0.269, 0.430, 0.698, and 0.861. The curves represent the
theoretical values calculated from Eq.~(\ref{uno}) with
$t_\mathrm{v}= 0.738$ s (the value derived from the fluid
properties at the mean temperature and at the critical density).
The topmost (dash-double-dotted) line is for equilibrium: $\Delta
T = 0$ ({\it i.e.}, $R = 0$). The remaining four curves are for
the values of $\Delta T/\Delta T_\mathrm{c}$ of the data sets, if
we adopt the Boussinesq estimate
\begin{equation}
R_\mathrm{OB} =  (\Delta T / \Delta T_\mathrm{c}) R_\mathrm{c}
\label{eq:ROB}
\end{equation}
with $R_\mathrm{c} = 1730$ (the value obtained from the Galerkin
approximation used in this paper~\cite{Manneville}) for the
Rayleigh number in Eq.~(\ref{unoa}). One sees that the
predictions, based on the Boussinesq approximation,  do not agree
very well with the experiment. However, both theory and experiment
reveal clearly a minimum of $\Gamma$ near $\tilde{q} \simeq 3$
which becomes more pronounced as $\Delta T$ approaches $\Delta
T_\mathrm{c}$. For the larger values of $\Delta T/\Delta
T_\mathrm{c}$ the experimental values of $\Gamma$ have a maximum
near $\tilde{q} = 5$. We expect that this is due to nonlinear
effects in the physical system which lead to second-harmonic
generation. This phenomenon is not contained in the theory.

We believe that the major disagreement between the measurements
and the calculation is due to symmetric non-OB effects discussed
in Sect.~\ref{sec:NOB}. Since there is no quantitative theory, we
proceeded empirically and explored the possibility  that non-OB
effects can be accommodated to a large extent by multiplying the
experimental $\Delta T/\Delta T_\mathrm{c}$ used to estimate $R$
by an adjustable scale factor. We introduced an adjustable
parameter $f_{R,k}$ which was allowed to be different for each
$\Delta T_k$ and used
\begin{equation}
R_k = f_{R,k} R_\mathrm{OB}
\end{equation}
for the Rayleigh number. In addition, we introduced a single
adjustable scale factor for all data sets which adjusted the
length scale of the experiment so as to yield a corrected 
wave number 
\begin{equation}
\tilde{q}_\mathrm{corr} = f_q \tilde{q}.
\end{equation}
to be used in the fit of the theory to the data. We expect $f_q$ to
compensate for experimental errors in the cell spacing 
and in the spacing between the pixels of the images, to
be within a few percent of unity,  and to be the same for the runs
at all $\Delta T$. Finally, we treated the vertical relaxation
time $t_\mathrm{v}$ as an adjustable parameter. We note that
$t_\mathrm{v} = d^2/D_T$ depends on the cell spacing, and thus any
error in the length scale will lead to an error in the time scale
$t_\mathrm{v}$. One also might expect $t_\mathrm{v}$ to depend on
$\Delta T$ because quadratic non-OB effects would be larger at
larger $\Delta T$; but it turned out that a single value for
$t_\mathrm{v}$ for the runs at all $\Delta T$ was sufficient to
describe all the data. We carried out a simultaneous least-squares
fit of Eq.~(\ref{eq:calR}) to a group of 11 data sets of $\Gamma$,
like those in Fig.~\ref{fig:Gamma0}, based on $\tau_0 = 0.500$ s
and $\tau_j = 0.200$ s. Each data set was for a different $\Delta
T_k$, and collectively they spanned the range $0.27 \leq \Delta
T_k/\Delta T_\mathrm{c} \leq 0.86$. A separate fit was done to the
second group of 9 available data sets based on $\tau_0$ = 0.500 s
and $\tau_j = 0.350$ s. For each group we simultaneously adjusted
$f_q, t_\mathrm{v}$, and the eleven or nine $f_{R,k}$. We obtained
the same result $f_q = 0.944 \pm 0.010$ from both groups. The fit
also gave $t_\mathrm{v} = 0.551 \pm 0.02$ ($0.565 \pm 0.029$) s
for the group based on $\tau_j = 0.200~(0.350)$ s. Qualitatively
consistent with the expected influence of the symmetric non-OB
effects, the fitted value of $t_\mathrm{v}$ is somewhat smaller
than the value 0.738 s estimated for $\rho=\rho_\mathrm{c}$ and
the experimental value $d = 34.3~\mu$m. As said above, part of
this difference is attributable to the error in $d$ indicated by
the result obtained for $f_q$. In Fig.~\ref{fig:Gamma1} we show
the results for the product $\tilde{\Gamma}=\Gamma t_v$ for the
four examples displayed in Fig.~\ref{fig:Gamma0} as a function of
$f_q\tilde{q}$, together with the corresponding predictions
generated by using the values of $f_{R,k}$ from the fit. Except
for the second-harmonic contribution at large $\tilde{q}$ and $\Delta T$,
the adjusted theory agrees quite well with the data.

The values obtained for $f_{R,k}$ are given in
Fig.~\ref{fig:paras}. The two groups ($\tau_j = 0.200$ s and
$\tau_j = 0.350$ s) agree very well with each other, showing that
consistent results are obtained with different exposure times.
Also shown is a fit to the data sets with $\tau_j = 0.200$ s in
which the individual $f_{R,k}$ were replaced by a quadratic
function which was forced to pass through $f_{R,k} = 1$ at $\Delta
T = \Delta T_\mathrm{c}$. This fit gave
\begin{equation}
f_{R}(\Delta T/\Delta T_\mathrm{c}) = 2.74 - 2.87 \frac{\Delta
T}{\Delta T_\mathrm{c}} + 1.13\left(\frac{\Delta T}{\Delta
T_\mathrm{c}}\right)^2\ .
\end{equation}
This two-parameter  representation of $f_{R}$ fits the data
equally well.

One sees that $f_{R}$ is largest at the smallest $\Delta T$,
indicating that the estimate $R = R_\mathrm{OB}$ (see
Eq.~(\ref{eq:ROB})) becomes worse as $\Delta T$ decreases. This is
to be expected because the  approximation for $f_{R}$ must
approach unity as $\Delta T$ approaches $\Delta T_\mathrm{c}$ in
order for $\Gamma(\tilde{q}_\mathrm{c}, \Delta T_\mathrm{c})$ to vanish.
As a simplest empirical attempt to include non-OB effects in the
comparison between experiment  and theory, we define, in analogy
to Eq.~(\ref{eq:ROB}), a non-Boussinesq Rayleigh number
\begin{equation}
R_\mathrm{NOB}(\Delta T) = \frac{\langle R(\rho) \rangle}{\langle
R(\rho)\rangle_\mathrm{c}}~R_\mathrm{c},
\end{equation}
where as before $R_\mathrm{c} = 1730$ and where the angular
bracket indicates an average over  the spanned temperature (and
thus density)  range along the isobar. The averaged critical
Rayleigh number $\langle R(\rho)\rangle_\mathrm{c}$ is equal to
$\langle R(\rho) \rangle$ for $\Delta T = \Delta T_\mathrm{c}$. It
turns out that $\langle R(\rho)\rangle_\mathrm{c} = 1120$ for our
experiment. This approximation corresponds to a re-defined
\begin{equation}
f_R(\Delta T) = \frac{\Delta T_\mathrm{c}}{\Delta T}~
\frac{\langle R(\rho) \rangle}{\langle R(\rho)
\rangle_\mathrm{c}}\ . \label{eq:f_R,NOB}
\end{equation}
We note that by definition $f_R(\Delta T_\mathrm{c}) = 1$ and thus
$R = 1730$ as it should be. Equation~(\ref{eq:f_R,NOB}) is plotted
in Fig.~(\ref{fig:paras}) as a dashed line. One sees that this
simplest non-OB model accounts very well for the experimental data
of $f_R$. Of course, it would be very helpful to have a proper
theory (rather than an empirical model) of this interesting
effect.

\section{Summary}
\label{sec:summary}

In this paper we have reported on experimental and theoretical
studies of the dynamics of thermal fluctuations below the onset of
Rayleigh-B\'enard convection in a thin horizontal fluid layer
bounded by two rigid walls and heated from below. Starting from
the fluctuating linearized Boussinesq equations, we derived
theoretical expressions for the dynamic structure factor and the
decay rates and amplitudes of the hydrodynamic modes that
characterize the dynamics of the fluctuations. The dynamic
structure factor is dominated by a slow mode with a decay rate
that vanishes as the Rayleigh number $R$ becomes equal to its
critical value $R_\mathrm{c}$ for the onset of convection.

We used the shadowgraph method to determine the ratio $\cal R$ of
shadowgraph structure factors obtained with different camera
exposure times. From the theoretical results for the dynamic
structure factor, we derived a relationship between this ratio and
the decay rates of the fluctuations. Using this result and the
experimental values of $\cal R$, we obtained experimental
decay-rate data for a wide range of temperature gradients below
the onset of Rayleigh-B\'enard convection in sulfur hexafluoride
near its critical point. Quantitative agreement between the
experimental decay rates and the theoretical prediction could be
obtained when allowance was made for some experimental uncertainty
in the small spacing between the plates and an empirical estimate
was employed for symmetric deviations from the Oberbeck-Boussinesq
approximation which are expected in a fluid with its mean density
on the critical isochore.\\

\section{ACKNOWLEDGEMENTS}

The research of J. Oh and G. Ahlers was supported by U.S. National
Science Foundation Grant DMR02-43336. G. Ahlers and J. M. Ortiz de
Z\'arate acknowledge support through a Grant under the Del Amo
Joint Program of the University of California and the Universidad
Complutense de Madrid. The research at the University of Maryland
was supported by the Chemical Sciences, Geosciences and
Biosciences Division of the Office of Basic Energy of the U.S.
Department of Energy under Grant No. DE-FG-02-95ER14509.

\begin{thebibliography}{10}

\bibitem{KirkpatrickEtAl}
T.~R. Kirkpatrick, E.~G.~D. Cohen, and J.~R. Dorfman, Phys. Rev. A {\bf 26},
  995  (1982).

\bibitem{LawSengers}
B.~M. Law and J.~V. Sengers, J. Stat. Phys. {\bf 57},  531  (1989).

\bibitem{SegreEtAl1}
P.~N. Segr\`{e}, R.~W. Gammon, J.~V. Sengers, and B.~M. Law, Phys. Rev. A {\bf
  45},  714  (1992).

\bibitem{miPRE}
{J. M.~Ortiz~de~Z\'{a}rate} and J.~V. Sengers, Phys. Rev. E {\bf 66},  036305
  (2002).

\bibitem{SegreSchmitzSengers}
P.~N. Segr\`{e}, R. Schmitz, and J.~V. Sengers, Physica A {\bf 195},  31
  (1993).

\bibitem{VailatiGiglio1}
A. Vailati and M. Giglio, Phys. Rev. Lett. {\bf 77},  1484  (1996).

\bibitem{Physica}
J.~M. {Ortiz de Z\'{a}rate}, R. {P\'{e}rez Cord\'{o}n}, and J.~V. Sengers,
  Physica A {\bf 291},  113  (2001).

\bibitem{Physica2}
J.~M. {Ortiz de Z\'{a}rate} and J.~V. Sengers, Physica A {\bf 300},  25
  (2001).

\bibitem{ZaitsevShliomis}
V. Zaitsev and M. Shliomis, Sov. Phys. JETP {\bf 32},  866  (1971).

\bibitem{SwiftHohenberg}
J.~B. Swift and P.~C. Hohenberg, Phys. Rev. A {\bf 15},  319  (1977).

\bibitem{HohenbergSwift}
P.~C. Hohenberg and J.~B. Swift, Phys. Rev. A {\bf 46},  4773  (1992).

\bibitem{WuEtAl}
M. Wu, G. Ahlers, and D.~S. Cannell, Phys. Rev. Lett. {\bf 75},  1743  (1995).

\bibitem{OA03}
J. Oh and G. Ahlers, Phys. Rev. Lett. {\bf 91},  094501  (2003).

\bibitem{LawEtAl}
B.~M. Law, P.~N. {Segr\`{e}}, R.~W. Gammon, and J.~V. Sengers, Phys. Rev. A
  {\bf 41},  816  (1990).

\bibitem{LekkerkerkerBoon}
H.~N.~W. Lekkerkerker and J.~P. Boon, Phys. Rev. A {\bf 10},  1355  (1974).

\bibitem{BoonEtAl}
J.~P. Boon, C. Allain, and P. Lallemand, Phys. Rev. Lett. {\bf 43},  199
  (1979).

\bibitem{LallemandAllain}
P. Lallemand and C. Allain, J. Phys. (Paris) {\bf 41},  1  (1980).

\bibitem{SchmitzCohen2}
R. Schmitz and E.~G.~D. Cohen, J. Stat. Phys. {\bf 40},  431  (1985).

\bibitem{KirkpatrickCohen}
T.~R. Kirkpatrick and E.~G.~D. Cohen, J. Stat. Phys. {\bf 33},  639  (1983).

\bibitem{AllainEtAl}
C. Allain, H.~Z. Cummins, and P. Lallemand, J. Phys. (Paris) Lett. {\bf 39},
  L473  (1978).

\bibitem{Sawada}
Y. Sawada, Phys. Lett. A {\bf 65},  5  (1978).

\bibitem{PedersenRiste}
A.~M. Pedersen and T. Riste, Z. Phys. B {\bf 37},  171  (1980).

\bibitem{OtnesRiste}
K. Otnes and T. Riste, Phys. Rev. Lett. {\bf 37},  1490  (1980).

\bibitem{BA77}
R. Behringer and G. Ahlers, Phys. Lett. {\bf 62A},  329  (1977).

\bibitem{WPDNB78}
J. Wesfreid, Y. Pomeau, M. Dubois, C. Normand, and P. Berg\'e, J. Phys.
  (France) {\bf 39},  725  (1978).

\bibitem{Ob79}
A. Oberbeck, Ann. Phys. Chem. {\bf 7},  271  (1879).

\bibitem{Bo03}
J. Boussinesq, {\em {Th\'eorie} Analytique de la Chaleur} (Gauthier-Villars,
  Paris, 1903), Vol.~2.

\bibitem{Ch61}
S. Chandrasekhar, {\em Hydrodynamic and Hydromagnetic Stability} (Oxford
  University Press, Oxford, 1961).

\bibitem{PlattenLegros}
J.~K. Platten and J.~C. Legros, {\em Convection in Liquids} (Springer, Berlin,
  1984).

\bibitem{Manneville}
P. Manneville, {\em Dissipative Structures and Weak Turbulence} (Academic
  Press, San Diego, 1990).

\bibitem{CrossHohenberg}
M.~C. Cross and P.~C. Hohenberg, Rev. Mod. Phys. {\bf 65},  851  (1993).

\bibitem{LandauLifshitz}
L.~D. Landau and E.~M. Lifshitz, {\em Fluid Mechanics} (Pergamon Press, London,
  1959).

\bibitem{DeBruynEtAl}
J.~R. de~Bruyn, E. Bodenschatz, S.~W. Morris, S.~P. Trainoff, Y. Hu, D.~S.
  Cannell, and G. Ahlers, Rev. Sci. Instrum. {\bf 67},  2043  (1996).

\bibitem{BrogioliEtAl}
D. Brogioli, A. Vailati, and M. Giglio, Phys. Rev. E {\bf 61},  R1  (2000).

\bibitem{MBCA96}
S.~W. Morris, E. Bodenschatz, D.~S. Cannell, and G. Ahlers, Physica D {\bf 97},
   164  (1996).

\bibitem{BPA00}
E. Bodenschatz, W. Pesch, and G. Ahlers, Annu. Rev. Fluid Mech. {\bf 32},  709
  (2000).

\bibitem{GrahamPleiner}
R. Graham and H. Pleiner, Phys. Fluids {\bf 18},  130  (1975).

\bibitem{SLB65}
A. Schluter, D. Lortz, and F.~H. Busse, J. Fluid Mech. {\bf 23},  129  (1965).

\bibitem{EPJ}
J.~M. {Ortiz de Z\'{a}rate} and L. {Mu\~{n}oz Redondo}, Euro. Phys. J. B {\bf
  21},  135  (2001).

\bibitem{TC02}
S.~P. Trainoff and D.~S. Cannell, Phys. Fluids {\bf 14},  1340  (2002).

\bibitem{GiglioNature}
A. Vailati and M. Giglio, Nature {\bf 390},  262  (1997).

\bibitem{VanBeijerenCohen}
H. van Beijeren and E.~G.~D. Cohen, J. Stat. Phys. {\bf 53},  77  (1988).

\bibitem{WS99}
A.~K. Wyczalkowska and J.~V. Sengers, J. Chem. Phys. {\bf 111},  1551  (1999).

\bibitem{HBT85}
J.~H.~B. Hoogland, H.~R. {van den B}erg, and N.~J. Trappeniers, Physica A {\bf
  134},  169  (1985).

\bibitem{SV89}
T. Strehlow and E. Vogel, Physica A {\bf 161},  101  (1989).

\bibitem{LK65}
J. Lis and P.~O. Kellard, Brit. J. Appl. Phys. {\bf 16},  1099  (1965).

\bibitem{SH73}
H.~L. Swinney and D.~L. Henry, Phys. Rev. A {\bf 8},  2586  (1973).

\bibitem{LimSwinney}
T.~K. Lim, H.~L. Swinney, K.~H. Langley, and T.~A. Kachnowski, Phys. Rev. Lett.
  {\bf 27},  1776  (1971).

\bibitem{Li73}
T.~K. Lim, Ph.D. thesis, Johns Hopkins University, 1973.

\bibitem{BT81}
V.~V. Burinskii and E.~E. Totski, High Temp. {\bf 19},  366  (1981).

\bibitem{KI85}
J. Kestin and N. Imaishi, Int. J. Thermophys. {\bf 6},  107  (1985).

\bibitem{Bu67a}
F.~H. Busse, J. Fluid Mech. {\bf 30},  625  (1967).

\end{thebibliography}

%
\end{document}